\documentclass[prb, superscriptaddress, twocolumn, amsmath, amssymb]{revtex4}

\setcitestyle{numbers,square}
\usepackage{amsmath,amssymb}
\usepackage{comment}
\usepackage{tabularx}
\usepackage{graphicx}
\usepackage{bmpsize}
\usepackage{bm}
\usepackage{xcolor}
\usepackage{amssymb}
\usepackage[version=3]{mhchem}
\usepackage{yhmath}
\usepackage{blkarray}
\DeclareMathAlphabet{\mathpzc}{OT1}{pzc}{m}{it}
\newcommand{\nn}{\nonumber}

\def\sgn{\mathrm {sgn}}

\definecolor{navy}{hsb}{0.66,1,0.6}

\begin{document}
\makeatletter
\renewcommand\@biblabel[1]{[#1]}
\makeatother

\preprint{APS/123-QED}

\title{Quantum higher-spin Hall insulators}

\author{Takuto Kawakami}
\email{tkawakm@fmu.ac.jp}
\affiliation{Center for Integrated Science and Humanities, Fukushima Medical University, Fukushima 960-1295, Japan}
\author{Igor Kuzmenko}
\affiliation{Department of Physics, Ben-Gurion University of the Negev, Beer-Sheva, 84105, Israel}
\author{Yshai Avishai}
\affiliation{Department of Physics, Ben-Gurion University of the Negev, Beer-Sheva, 84105, Israel}
\affiliation{Yukawa Institute for Theoretical Physics, Kyoto University, Kyoto 606-8502, Japan}
\author{Yigal Meir}
\affiliation{Department of Physics, Ben-Gurion University of the Negev, Beer-Sheva, 84105, Israel}
\author{Masatoshi Sato}
\affiliation{Yukawa Institute for Theoretical Physics, Kyoto University, Kyoto 606-8502, Japan}

\date{\today}

\begin{abstract}

We develop a theory of quantum spin Hall insulators with arbitrary spin $J$.
Our analysis demonstrates that such systems support $J+\tfrac{1}{2}$ pairs of helical edge modes
protected by nontrivial mirror Chern numbers.
We establish that the corresponding edge theory is described by a
generalized Dirac fermion with higher-order dispersion.
These modes produce unique transport responses that are non-linear with voltage.
An in-plane magnetic field opens a mass gap in the edge spectrum, and
magnetic domain walls host $(J+\tfrac{1}{2})$-fold degenerate bound states
characterized by nontrivial winding numbers.
Our results extend quantum spin Hall physics to higher-spin systems
and suggest possible realizations in ultracold atomic gases.

\end{abstract}


\maketitle

Recently, higher-spin fermionic systems have emerged as a fertile setting for realizing novel topological phases of matter~\cite{bergman2006, hsieh2014, brydon2016, agterberg2017, timm2017, ghrashi2017, savary2017,yang2017,venderbos2018, brydon2018, yu2018, kobayashi2019,roy2019, menke2019, kobayashi2020, tchoumakov2020,szabo2021,  dutta2021, kim2021, kim2021jpsj,timm2021,timm2021distortional, kim2022,kobayashi2022, sato2024,ohashi2024,mori2025, oudah2016, kawakami2018, ikeda2019, chenfang2019, ikeda2020, boettcher2020, link2020, kobayashi2021, mandal2023,bhattacharya2023,hasanreview, qireview, tanakareview, yandoreview, andofureview, mizushimareview,msatoreview, msatoandoreview,chiureveiw}. 
In solid-state systems, although electrons carry intrinsic spin-$1/2$, 
spin-orbit coupling mixes orbital and spin degrees of freedom, 
giving rise to quasiparticles with effective total angular momentum $J=3/2$. 
A prominent example occurs in Half-Heusler compounds described by the Luttinger–Kohn model
~\cite{brydon2016, agterberg2017, timm2017, ghrashi2017, savary2017,yang2017,venderbos2018, brydon2018, yu2018, kobayashi2019,roy2019, menke2019, kobayashi2020, tchoumakov2020,szabo2021,  dutta2021, kim2021, kim2021jpsj,timm2021,timm2021distortional, kim2022,kobayashi2022,sato2024,ohashi2024,mori2025}, 
as well as in antiperovskites exhibiting band inversion of $J=3/2$ electrons
~\cite{hsieh2014,kariyado2011,kariyado2012,oudah2016,kawakami2018,ikeda2019,chenfang2019, ikeda2020}. 
These systems have been predicted to host a variety of unconventional topological phases, including topological insulators and high-winding topological superconductors.

Ultracold atomic gases provide a complementary platform for exploring topological phenomena, 
owing to their high degree of tunability via external fields and optical techniques~\cite{cooperreview, jaksch2003, osterloh2005,raghu2008, umucalilar2008,shao2008,wu2008, msato2009, msato2010,goldman2010,jiang2011,goldman2013, beeler2013, kennedy2013, aidelsburger2013, cao2014, duca2015, rli2016, grusdt2017, zheng2019, zhangreview2018, jyzhang2023}. 
Recent experimental advances have enabled the realization of synthetic spin-orbit coupling
~\cite{dalibard2011,zhai2015, ruseckas2005,juzeliunas2010,lin2011,campbell2011,anderson2012,wu2016,lhuang2016}
through Raman coupling schemes, which is a key ingredient for engineering topological phases.
As a result, a wide range of phases—including the anomalous quantum Hall effect, the quantum spin Hall effect, and topological superfluidity—have been proposed in cold-atom settings~\cite{umucalilar2008,shao2008,wu2008, goldman2013,goldman2010, beeler2013, kennedy2013, grusdt2017,msato2009,msato2010,jiang2011,mizushima2013,cao2014}.

A particularly attractive feature of ultracold atomic gases is the natural realization of high-spin degrees of freedom.
For example, isotopes such as $^6$Li and $^2$He possess hyperfine spin $3/2$, 
while the ground state of $^{40}$K has spin $9/2$. 
Recent theoretical work has predicted topological insulating phases in two-dimensional
spin-orbit-coupled systems of spin-$3/2$ atoms~\cite{kuzmenko2018}. 
These studies introduced spin-$3/2$ analogs of helical edge states characterized by mirror Chern numbers and proposed topological phase transitions driven by out-of-plane Zeeman fields. 
Nevertheless, a systematic understanding of quantum spin Hall physics for general spin $J$ systems remains largely unexplored.

In conventional spin-$1/2$ quantum spin Hall insulators (QSHIs), 
magnetic domain walls at the edge host localized bound states
~\cite{jackiw, qi2008, gao2009, kharitonov2012,qireview, dalum2021}. 
The edge modes of a spin-$1/2$ QSHI form one-dimensional massless Dirac fermions, 
and Zeeman or exchange fields open a mass gap in the spectrum. 
When the magnetic field changes sign across a domain wall, the mass term reverses sign, producing a Jackiw–Rebbi bound state localized at the interface~\cite{jackiw}. 
How this mechanism generalizes to quantum spin Hall systems with higher spin $J$ remains an open question.

\begin{figure*}[t]
\begin{center}
	\includegraphics[width=170mm]{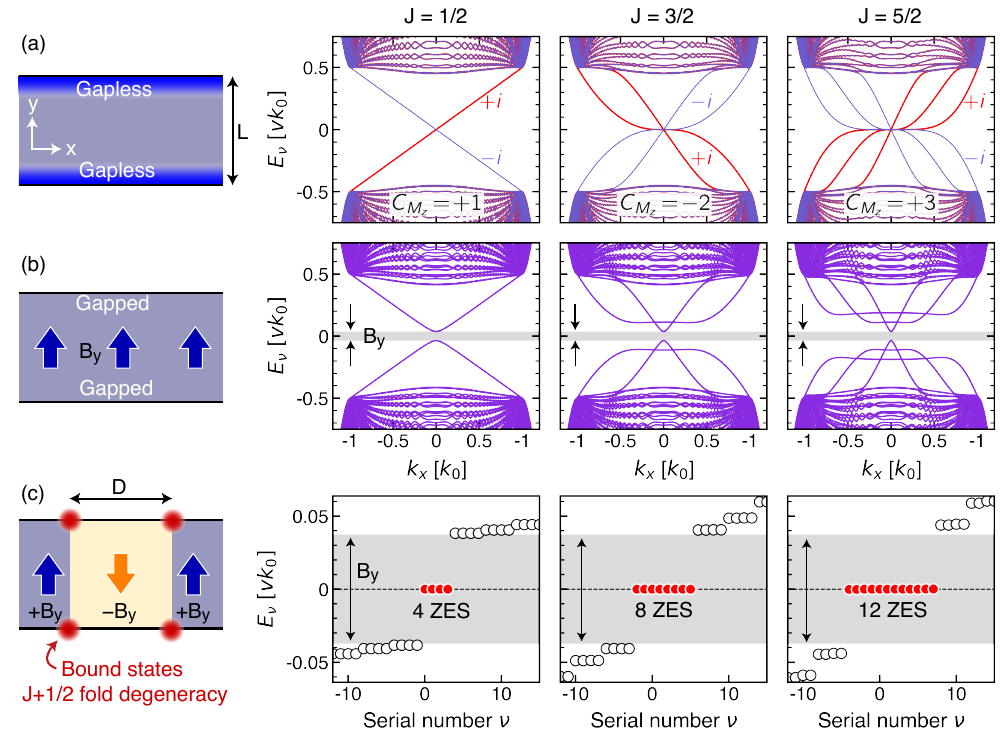}
	\caption{Geometry (left) and energy spectrum (right three columns for spin 1/2, 3/2, and 5/2) of the ribbon structure under several different in-plane magnetic field configurations. 
	Panels (a) corresponds to the absence of a magnetic field ($B_y=0$), where the red and blue curves represent 
	states characterized by eigenvalues $+i$ and $-i$ of the mirror reflection operator $M_z$, respectively.
	Panels (b) depicts a uniform magnetic field ($B_y=0.075 vk_0$), and panels 
	(c) displays a domain wall with $B_y=0.075vk_0$. 
	In (a) edge states localized at $y=L$ are omitted for clarity. 	
    The wave vector is measured in units of $k_0=\sqrt{\epsilon_0}$ with $\hbar=1$ and $2M=1$, and we set $k_0/v=2.5$. 
	The width of the ribbon is $L=100 k_0^{-1}$. The domain walls separation is $D=200 k_0^{-1}$. 
	In panels, (c) the periodic boundary condition with length $2D$ is applied along $x$-direction.
	Notably, zero-energy states emerge at the four points highlighted in the left panel. 
	Each of these points exhibits single, double, and triple degenerate zero modes corresponding to spin values of $J=1/2$, $3/2$, and $5/2$ respectively.
	}\label{fig:spectrum}
\end{center}
\end{figure*}

In this Letter we investigate quantum spin Hall insulators formed by fermions with general spin $J$, 
with possible realizations in ultracold atomic gases. 
Our numerical analysis reveals that spin-$J$ systems host $J+\tfrac12$ pairs of gapless helical edge modes, 
consistent with bulk–edge correspondence associated with mirror Chern numbers. 
We derive the effective edge theory and show that it is described by
generalized Dirac fermions with higher-order dispersion $k^{|2m|}$ 
($m= J, J-1,\ldots,-J$), 
where $m$ denotes the in-plane magnetic quantum number.
The higher-order dispersion leads to nonlinear transverse transport in addition to the quantized $(J+\tfrac12)$ spin Hall response.

Finally, we demonstrate that applying an in-plane Zeeman field gaps the edge spectrum.
At domain walls between regions with opposite Zeeman fields, $(J+\tfrac12)$-fold degenerate bound states appear.
The number of domain-wall states is determined by
the winding number of the effective edge Hamiltonian.

{\it Quantum higher-spin Hall insulator. ---}
We consider the Hamiltonian of cold atoms subjected to spin-orbit coupling for general spin $J$~\cite{kuzmenko2018}
\begin{align}\label{eq:hbulk}
H=  \epsilon_k \sigma_z +  \hbar v \bm{k}\cdot \bm{J} \sigma_x
\end{align}
where $\sigma_i$ is the Pauli matrix acting on the orbital (or bands), and
the vector $\bm{J}=(J_x,J_y)$ is the spin-$J$ matrix with 
magnitude $J=N-\frac{1}{2}$ ($N=1,2,\cdots$).
The explicit matrix representation of these operators is provided in the Appendix.
Then, $\bm{k}=(k_x,k_y)$ is the two-dimensional momentum, and
$\epsilon_k=(\frac{\hbar^2}{2M}\bm{k}^2 - \epsilon_0 )$.
Here and in the following, we set $\hbar = 1$ and $2M = 1$ for simplicity, unless explicitly mentioned otherwise.
For $\epsilon_0>0$, the first term of Eq.~\eqref{eq:hbulk} gives the band inversion at $\bm{k}=0$, implying a topological phase.
The second term accounts for the spin-orbit coupling 
resulting in an energy gap within the bulk.

To explore the topological properties of the system, 
we calculate the energy spectrum in the stripe geometry,
with finite width along the $y$-direction, while
maintaining translational symmetry along the $x$-direction. 
By replacing $k_y$ with $-i\partial_y$ and treating $k_x$ as a well-defined quantum number, 
we obtain the energy spectrum, as shown in Fig.~\ref{fig:spectrum}(a). 
The present system is
an extension of the QSHI to encompass the spin-$J$. 
For $J = 1/2$, the system is an ordinary QSHI supporting helical edge states with linear dispersion. 
In contrast, for general spin $J$, there appear $J + 1/2$ pairs of gapless edge states with higher-order dispersion, demonstrating unique topological attributes of the system. 


{\it Mirror Chern numbers.---}
The gapless edge states displayed in Fig.~\ref{fig:spectrum}(a)
find their topological origin in the mirror Chern number. 
The Hamiltonian in Eq.~\eqref{eq:hbulk} exhibits mirror reflection symmetry
\begin{align}\label{eq:msym}
	M_z H M_z^{\dag} = H \quad \hbox{with } M_z=\sigma_z e^{iJ_z\pi}.
\end{align}
Through the application of a unitary matrix $U=\tfrac{1}{2}[(1-ie^{iJ_z\pi}) + (1+ie^{iJ_z\pi})\sigma_x]$
that diagonalizes the mirror operator $M_z$, 
the Hamiltonian is divided into two sectors $UHU^\dag=\mathrm{diag}(H_{+},H_{-})$ with 
\begin{equation}\label{eq:hpm}
	H_{\pm} = \mp ie^{iJ_z \pi} \epsilon_k + v \bm{k}\cdot\bm{J}.
\end{equation}
These sectors correspond to mirror eigenvalues $\pm i$ of $M_z$ ($\hbar=1$ units).
With the eigenstates $|\psi_{\nu}^{\pm}(\bm{k})\rangle$ of $H_{\pm}$ as the basis, 
the Chern number within each mirror sector is defined as 
\begin{align}
	C_{\pm}\!=\!\sum_{\nu\in \mathrm{occ}} \int \frac{dk}{2\pi} \left(\bm{\nabla}_{\bm{k} }\times \bm{a}_{\pm, \nu}\right)_z,
\end{align}
with the Berry connection $\bm{a}_{\pm, \nu} = i\langle \psi_{\nu}^\pm |\bm{\nabla}_{\bm{k}} \psi_{\nu}^\pm  \rangle$.
The mirror Chern number, representing the difference between the Chern numbers of the two sectors $C_{M_z}=\frac{1}{2}(C_{+}-C_{-})$ 
serves as a topological invariant characterizing the present system.

For general spin $J$, the mirror Chern number is given by
\begin{equation}\label{eq:cmmain}
	C_{M_z} \!=\! 1 - 3 + 5 - \cdots+\!(-1)^{J-\frac{1}{2}}2J = (-1)^{J-\frac{1}{2}} \!\left(J\!+\!\frac{1}{2}\right)\!,
\end{equation}
as shown in Appendix.
Correspondingly, Fig.~\ref{fig:spectrum}(a) shows that the number of the time-reversal pairs of edge states is $J+1/2$, directly reflecting $|C_{M_z}|$. A change in the sign of $C_{M_z}$ induces a sign reversal of the edge-state velocity in the mirror $\pm i$ sectors, highlighting the underlying topological origin.

{\it Edge states with higher-order dispersion.---}
Now, we derive the effective 1D model for the edge states.
Consider a semi-infinite system in the region $y\geq 0$, 
and translational symmetry along $x$-direction, 
thus rendering the wave number $k_x$ to be a well-defined quantum number.
Introducing $k_0 = \sqrt{\epsilon_0}$, the $k_x^2$ term in Eq.~\eqref{eq:hbulk} can be neglected for sufficiently small \(k_x \ll k_0\). The Hamiltonian in Eq.~\eqref{eq:hbulk} can then be written as $H\sim H_0 + (k_x/k_0)H_1 $
with
\begin{gather}
H_0 = \left(-\partial_y^2-k_0^2\right) \sigma_z - iv \partial_y J_y \sigma_x, \label{eq:h0} \\
H_1 = v k_0 J_x \sigma_x, 
\end{gather}
Below, we treat $H_1$ as a perturbation.

Let us first consider the unperturbed Schr\"odinger equation
$H_0 | \nu \rangle = E_{\nu}|\nu \rangle$.
Since $H_0$ commutes with $J_y$, we label the eigenstates by the magnetic quantum number associated with $J_y$:
\begin{equation}\label{eq:my}
m=J,J-1,\cdots -J.
\end{equation}
Far from the edge, 
the eigenstates reduce to plane waves with energy
\begin{align}
	E_\nu  
        = \eta\sqrt{\epsilon^2_{k_y}+(v m k_y)^2}, \quad \epsilon_{k_y} = k_y^2-k_0^2,
\end{align}
where $k_y$ denotes the wave number along $y$-direction, and $\eta=\pm$ is the particle-hole index.
Hereafter, $\nu$ denotes the collective index $(k_y, m, \eta$).
The corresponding wave function reads
\begin{equation}~\label{eq:pw}
	| \nu \rangle = 
	\frac{e^{ik_y y}}{\sqrt{L \mathcal{N}_{\mathrm{b}} }} 
	\left(\begin{array}{c}
		vm k_y \\
		E_{\nu} -  \epsilon_{k_y}
	\end{array}\right)
	|m \rangle.
\end{equation}
Here, $\mathcal{N}_{\mathrm{b}}=2E_{\nu}(E_{\nu}-\epsilon_{k_y})$
is the normalization constant. 
$L$ denotes the system size corresponding to width in the stripe geometry. 
$|m \rangle$ represents the eigenvector of
$J_y$ with $m$ in Eq.~\eqref{eq:my}.

Near the edge, in addition to the plane-wave states in Eq.~\eqref{eq:pw}, evanescent modes arise for complex $k_y$ with $\mathrm{Im}(k_y) > 0$.
In particular, for
\begin{equation}\label{eq:cmpk}
	k_y= \pm K_{m} + i\kappa_{m},
\end{equation}
with $K_{m}\!=\! \sqrt{k_0^2 \!-\! (m v/2)^2}$ and $\kappa_{m}\!=\!|m v|/2$, we obtain two zero-energy solutions for each spin component $m$. 
A physical edge state is constructed as a linear combination of these solutions. 
We impose a vanishing wave function at the boundary $y=0$. The resulting wave function is given by
\begin{align}\label{eq:edge}
	| \chi_{m}\rangle = \frac{e^{-\kappa_{m} y}}{\sqrt{\mathcal{N}_0}} 
        \sin(K_{m} y)\left(\begin{array}{c}
		1 \\
		-i \sgn(m)
	\end{array}\right) | m \rangle
\end{align}
The normalization constant is $\mathcal{N}_0=\frac{1}{2}(\frac{1}{\kappa_m} - \frac{\kappa_m}{k_0^2})$.
For a general spin-$J$, the total number of zero energy states is $2J+1$, 
consistent with the numerical results at $k_x=0$ shown in Fig.~\ref{fig:spectrum}(a).


For finite $k_x$, 
the effective Hamiltonian for the subspace $s=\{ |\chi_{|m|}\rangle,|\chi_{-|m|}\rangle \}$ is obtained within $|2m|$-th order perturbation theory:
\begin{align}\label{eq:heff}
	\mathcal{H}_{|m|}=
	h_{|m|}\left(\begin{array}{cc}
		0 &  (-i k_x)^{|2m|} \\
		\left(i k_x\right)^{|2m|} & 0 
	\end{array}\right),
\end{align}
where the coefficient is given by 
\begin{equation}\label{eq:coeff}
        h_{|m|} = (ik_0)^{-|2m|}\langle \chi_{-|m|}|H_1 (H_0 ^{-1}H_1)^{|2m|-1}|\chi_{|m|}\rangle 
\end{equation}
It is evaluated numerically as a function of the material parameter  $k_0/v$ (see Fig.~\ref{fig:integ}). A detailed derivation is provided in the Appendix.
Each sector of the effective Hamiltonian in Eq.~\eqref{eq:heff}
has eigenvalues $E\propto\pm k_x^{|2m|}$. 
The edge states in the higher-spin quantum spin Hall insulator are thus one-dimensional massless Dirac fermions with nonlinear dispersion.


\begin{figure}[b]
\centering
	\includegraphics[width=85mm]{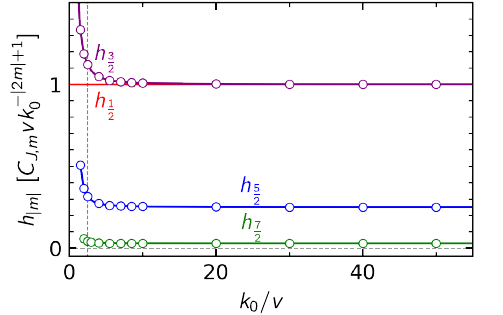}
	\caption{Numerically evaluated $h_{|m|}$ in Eq.~\eqref{eq:coeff} in the unit of $C_{J,m}v k_0$, where the $J$-depending coefficient is given by $C_{J,m}=\prod_{m'=-m+1}^{m} \frac{1}{2}\sqrt{(J+m')(J-m'+1)}$.
	}\label{fig:integ}
\end{figure}

{\it Nonlinear longitudinal conductance.---}
The helical edge states in the conventional quantum spin Hall insulators give a quantized linear longitudinal conductance, which has been experimentally observed in HgTe quantum well~\cite{konig2007, roth2009} and InAs/GaSb bilayers~\cite{knez2011,du2015}. 
Here, we examine longitudinal transport associated with the edge state of the quantum higher-spin Hall insulator.

In an ideal clan situation, ballistic transport leads to a quantized conductance of $Ne^2/h$ where $N$ is the number of the edge states. 
However, in a real system, diffusive scattering may occur along the edge, and the transport is described by the Boltzmann equation. 
In the latter case, with weak disorder, the edge modes of the present system exhibit nonlinear longitudinal conductance due to higher-order dispersion, as shown below.


Within the relaxation time approximation, 
the Boltzmann equation under the potential gradient $U(x)=\mathcal{E} x$ along the edge
is given by~\cite{ashcroftmermin, kawabata2022} 
\begin{equation}\label{eq:boltzman}
	\frac{\mathcal{E}}{\hbar}\frac{d f_{\alpha}}{d k_x} = -\frac{1}{\tau}\Big(f_{\alpha}-f_0(E_{\alpha})\Big)
\end{equation}
where $f_0(E) = (1+e^{\beta E})^{-1}$ is the Fermi-Dirac distribution, 
$f_{\alpha}$ is the non-equilibrium distribution function with a collective index for $k_x$, $|m|$, $E_{\alpha}=\eta h_{|m|} k_x^{|2m|}$ ($\eta=\pm$), and we restore $\hbar$ for clarity.
Then, the current density is given by
\begin{align}
	j_{\pm,m}\!=\! \int \frac{d k_x}{2\pi} v_x (f_\alpha\!-\!f_0(E_{\alpha}))
        \!=\!\frac{|2m|!}{2\pi\hbar}
        h_{|m|}\left(\frac{\mathcal{E}\tau }
        {\hbar}\right)^{|2m|}.\nn
\end{align}
where $v_x=\hbar^{-1}d E_{\alpha}/d k_x$ is the group velocity. (For a detailed derivation, see Appendix.)
Thus, we obtain the $n$-th order non-linear conductivity, defined by $j=\sum_{\pm,m_y}j_{\pm,m_y}=\sum_{n}\sigma^n\mathcal{E}^n$,
\begin{equation}\label{eq:sigma}
    \sigma^{(n)} =  \frac{\tau^{n} n!}{\pi {\hbar}^{n+1}}  h_{n/2}.
\end{equation}
Therefore, the non-parabolic $k^{|2 m|}$ dispersion of the higher spin edge states results in an $|2m|$-th order non-linear conductivity as a leading non-ballistic contribution.

{\it Edge state engineering.---}
We can control the edge states by applying an in-plane magnetic field 
\begin{equation}\label{eq:hb}
H_{\mathrm{B}} = H + B_y J_y.
\end{equation}
The uniform in-plane magnetic field along the $y$ direction
breaks both the mirror reflection symmetry $M_z$ and time-reversal symmetry,  
opening an energy gap at the edge for all $J$,
as demonstrated in Fig.~\ref{fig:spectrum}(b). 
Furthermore, when domains with opposing magnetic fields are present,  
as depicted in Fig.~\ref{fig:spectrum}(c), 
localized states emerge at the domain walls within the gap.
Notably, the spin $J$ 
and the number $N$ of the domain-wall bound states has the following intimate relation, 
\begin{align}
	N=J+\frac{1}{2}.
\end{align}
For spin 1/2, the bound state $N=1$ corresponds to 
the so-called Jackiw-Rebbi bound states in quantum spin Hall insulators~\cite{jackiw, qi2008, gao2009, kharitonov2012,qireview}.
The above results generalize it to the quantum higher-spin Hall insulator.

We now identify the topological number characterizing the domain-wall bound states. 
The key symmetry here is the chiral symmetry of the original Hamiltonian Eq.~\eqref{eq:hb}:
\begin{equation}
	\Gamma H_{\mathrm{B}} \Gamma^{\dag} = -H_{\mathrm{B}} \quad\hbox{with } \Gamma = i\sigma_y M_z
\end{equation}
On the other hand, the effective Hamiltonian in the $m_y$ sector with the magnetic field applied acting in the subspace $\left\{|\chi_{|m|}\rangle, |\chi_{-|m|}\rangle \right\}$ is described as
\begin{equation}
	\mathcal{H}_{|m|,\mathrm{B}} = 
	\mathcal{H}_{|m|} + 
	B_y |m| \left( \begin{array}{cc}
		1 & 0 \\
		0 & -1
	\end{array}\right)
\end{equation}
where ${\cal H}_{|m_y|}$ is given in Eq.(\ref{eq:heff}). 
The matrix representation of $\Gamma$ in this subspace is
\begin{equation}
	\gamma_{|m|} = 
	(-1)^{J-|m|}\left(\begin{array}{cc}
		0&  1 \\
		1&  0\\
	\end{array}\right), 
\end{equation}
which gives a chiral symmetry $\gamma_{|m|}\mathcal{H}_{|m|,B}\gamma_{|m|}=-\mathcal{H}_{|m|,B}$.
In the presence of the chiral symmetry, we can define the winding number~\cite{sato2011} in each $|m_y|$ sector 
\begin{align}
	w(|m_y|, B_y) &\equiv \int \frac{d k_x}{4\pi i} \mathrm{tr}\left( \gamma_{|m_y|}  \mathcal{H}_{|m_y|,\mathrm{B}}^{-1}\partial_{k_y} \mathcal{H}_{|m_y|,\mathrm{B}} \right) \nn\\ 
	&= \frac{1}{2}(-1)^{J+1}\mathrm{sgn}(h_{|m_y|} B_y).
\end{align}
Notably, the winding number in each sector does not depend on spin component $m$, because $h_{|m|}>0$ as shown in Fig.~\ref{fig:integ}.
Therefore, the difference in the total winding number $W(B_y)=\sum_{|m|} w(|m|, B_y)$ between the domains with $B_y$ and $-B_y$ for spin $J$ is
\begin{align}
	W(B_y)-W(-B_y) = (-1)^{J+\frac{1}{2}} \mathrm{sgn}(B_y)\left(J+\frac{1}{2} \right) \label{eq:winding}.
\end{align}
The winding number Eq.~\eqref{eq:winding} clearly ensures the presence of $J+\frac{1}{2}$ bound states consistently with the numerical results shown in Fig.~\ref{fig:spectrum}(c).

{\it Conclusion---}
We have investigated the topological properties of 
quantum spin Hall phases realized in systems of fermions with general spin $J$.
Our results extend the concept of quantum spin Hall insulators to higher-spin systems,
with possible realizations ranging from ultracold atomic gases to
solid-state platforms hosting high-spin quasiparticles,
such as antiperovskite topological materials.
Our system hosts $J+\tfrac12$ pairs of gapless helical edge modes,
protected by the mirror Chern number 
$C_{M_z}=(-1)^{J-1/2}(J+\tfrac12)$.
Each branch of the edge spectrum is described by a generalized one-dimensional Dirac fermion
with a higher-order dispersion determined by the in-plane spin component.
The higher-order dispersion leads to a characteristic nonlinear non-balistic transport.

When a magnetic domain wall separates regions with opposite in-plane magnetic fields, the interface hosts topologically protected $(J+\tfrac12)$-fold degenerate bound states.
From the standard counting and charge neutrality arguments, the $(J+\tfrac12)$-degeneracy results in a series of non-trivial charges that the interface between the domain wall and the boundary possesses,  $Q=-(J+\tfrac12)e/2, -(J+\tfrac12)e/2+1, -(J+\tfrac12)e/2+2, 
\dots, (J+\frac12)e/2$.
Thus, the higher-spin Hall insulator naturally admits higher-charge solutions, of which the charge takes either integer or half-integer values, depending on $J$. 
Hence, the higher-spin Hall insulator could naturally realize a quantum device with a higher charge.


\begin{acknowledgments}
This work was supported by JSPS KAKENHI
Grant Nos. JP24K06921, JP24K00569, JP25H01250, 
and JSPS and ISF under the Japan-Israel Research Cooperative Program.
The numerical calculations were carried out on Yukawa-21 at YITP in Kyoto University.
\end{acknowledgments}


\appendix
\section{Matrix Representation of spin J} \label{sec:spinmat}
The angular momentum is defined as a Hermitian operator with the commutation relation
\begin{align}\label{eq:comm}
	[\hat{J}_i,\hat{J}_j]=i \epsilon_{ijk}\hat{J}_k.
\end{align}
The the simultaneous eigenvector of the $\bm{J}^2$ and $J_z$ satisfies
\begin{align}
	\hat{\bm{J}}^2| J, m_z \rangle &= J(J+1)| J, m_z\rangle, \\ 
	J_z| J, m_z \rangle & = m_z | J, m_z\rangle.
\end{align}
where $J$ is the magnitude of the spin and $m_z=J, J-1,\cdots,-J$ is the $z$-compoent.
Using the commutation relation \eqref{eq:comm},
the matrix element of $J_x$ and $J_y$ is given by
\begin{align}
	\langle J, m_z | \hat{J}_x | J, m'_z \rangle &=
		\frac{1}{2} \sqrt{(J\pm m_z)(J\mp m_z+1)} \delta_{m_z,m'_z\pm 1} \\
	\langle J, m_z | \hat{J}_y | J, m'_z \rangle &= 
		\mp \frac{i}{2} \sqrt{(J\pm m_z)(J\mp m_z+1)} \delta_{m_z,m'_z\pm 1} \\
	\langle J, m_z | \hat{J}_z | J, m'_z \rangle &= 
		 m_z \delta_{m_z,m'_z}
\end{align}

For spin $J=1/2$ the matrix representations are 
\begin{align}
	J_x &= \frac{1}{2}
	\left(\begin{array}{cc}
	0 & 1 \\
	1 & 0 
	\end{array}\right) \\
	J_y &= \frac{1}{2}
	\left(\begin{array}{cc}
	0 & -i \\
	i & 0 
	\end{array}\right) \\
	J_z &= \frac{1}{2}
	\left(\begin{array}{cc}
	1 & 0 \\
	0 & -1 
	\end{array}\right)
\end{align}
and corresponding to the $2\times2$ Pauli matrices.
For spin $J=3/2$, we have 
\begin{align}
	J_x &= 
	\left(\begin{array}{cccc}
	0 & \tfrac{\sqrt{3}}{2} & 0 & 0 \\
	\tfrac{\sqrt{3}}{2} & 0 & 1 & 0 \\
	0 & 1 & 0 & \tfrac{\sqrt{3}}{2} \\
	0 & 0 & \tfrac{\sqrt{3}}{2} & 0\\
	\end{array}\right) \\
	J_y &= 
	\left(\begin{array}{cccc}
	0 & -i\tfrac{\sqrt{3}}{2} & 0 & 0 \\
	i\tfrac{\sqrt{3}}{2} & 0 & -i & 0 \\
	0 & i & 0 & -i\tfrac{\sqrt{3}}{2} \\
	0 & 0 & i\tfrac{\sqrt{3}}{2} & 0\\
	\end{array}\right) \\
	J_z &= 
	\left(\begin{array}{cccc}
	\tfrac{3}{2} & 0 & 0 & 0 \\
	0 & \tfrac{1}{2}& 0 & 0\\
	0 & 0&-\tfrac{1}{2} & 0\\
	0 & 0 & 0&-\tfrac{3}{2}\\
	\end{array}\right).
\end{align}
For spin $J=5/2$, we have 
\begin{align}
	J_x &= 
	\left(\begin{array}{cccccc}
	0 & \tfrac{\sqrt{5}}{2} & 0 & 0 & 0 & 0\\
	\tfrac{\sqrt{5}}{2} & 0 & \sqrt{2} & 0 & 0 & 0\\
	0 & \sqrt{2} & 0 & \tfrac{3}{2} & 0 & 0 \\
	0 & 0 & \tfrac{3}{2} & 0 & \sqrt{2} & 0\\
	0 & 0 & 0 & \sqrt{2} & 0 & \tfrac{\sqrt{5}}{2} \\
	0 & 0 & 0 & 0 &  \tfrac{\sqrt{5}}{2} & 0
	\end{array}\right) \\
	J_y &= 
	\left(\begin{array}{cccccc}
	0 & -i\tfrac{\sqrt{5}}{2} & 0 & 0 & 0 & 0\\
	i\tfrac{\sqrt{5}}{2} & 0 & -i\sqrt{2} & 0 & 0 & 0\\
	0 & i\sqrt{2} & 0 & -i\tfrac{3}{2} & 0 & 0 \\
	0 & 0 & i\tfrac{3}{2} & 0 & -i\sqrt{2} & 0\\
	0 & 0 & 0 & i\sqrt{2} & 0 & -i\tfrac{\sqrt{5}}{2} \\
	0 & 0 & 0 & 0 &  i\tfrac{\sqrt{5}}{2} & 0
	\end{array}\right) \\
	J_z &= 
	\left(\begin{array}{cccccc}
	\tfrac{5}{2} &0 &  0 & 0 & 0 & 0\\
	0 & \tfrac{3}{2} & 0 & 0 & 0 & 0\\
	0 & 0 & \tfrac{1}{2}& 0  & 0 & 0\\
	0 & 0 & 0&-\tfrac{1}{2}  & 0 & 0\\
	0 & 0 & 0 & 0&-\tfrac{3}{2}  & 0\\
	0 & 0 & 0 & 0& 0& -\tfrac{5}{2}
	\end{array}\right).
\end{align}
For more larger $J$, we can derive them following the same rule. In the following appendix and main text, $J$ in the ket is omitted and refered to as $|J,m_z\rangle=|m_z\rangle$. Additionally, we use $i=x$, $y$, $z$ as the spin quantization axis and denote it as $|m_i\rangle$.

\begin{figure}[b]
\begin{center}
	\includegraphics[width=85mm]{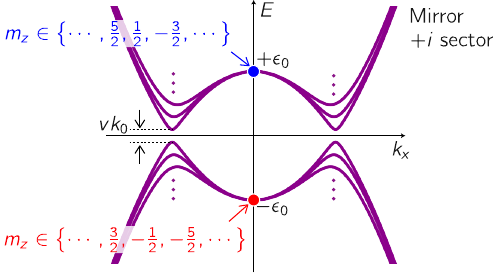}
	\caption{Schematic plot of dispersion relation for the bulk in mirror $+i$ sector.
	}\label{fig:bulk}
\end{center}
\end{figure}

\section{Mirror Chern Number for Spin-J} \label{sec:mchapp}
We compute the mirror Chern number for the system with general spin-$J$ as described Eq.~\eqref{eq:cmmain}.
In particular, we narrow our focus to the the sector associated with the eigenvalue $+i$ of the mirror operator in Eq.~\eqref{eq:hpm},
as governed by the Schr\"odinger equation
\begin{align}
	H_{+} | \psi_\nu^{+}(\bm{k}) \rangle = E_\nu(k) | \psi_\nu^{+}(\bm{k}) \rangle,
\end{align}
The dispersion relation $E_\nu(k)$ is plotted in in Fig.~\ref{fig:bulk}. 
It is important to note that the mirror $-i$ sector shares the same 
energy levels as the $+i$ sector, 
while it carries angular momentum along  $z$ direction with opposite signs.

Our goal is to calculate the Chern number specifically for the mirror $+i$ sector:
\begin{align}
	C_{+}=\sum_{\nu\in \mathrm{occ}} \int dk \left(\bm{\nabla}_{\bm{k} }\times \bm{a}_{+,\nu}\right)_z
\end{align}
where the Berry connection is defined as
\begin{align}
	\bm{a}_{+,\nu} = i\langle \psi_\nu^{+}| \bm{\nabla}_{\bm{k}} \psi_{\nu}^{+} \rangle 
\end{align}

\begin{figure}[t]
\begin{center}
	\includegraphics[width=85mm]{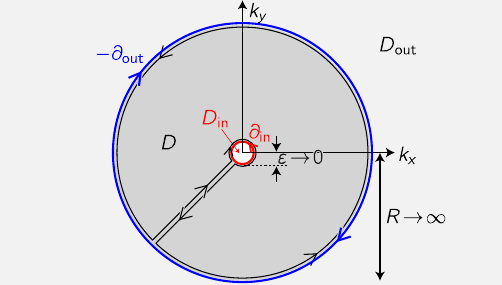}
	\caption{Patch of domains in momentum space to calculate the Chern number. 
	}\label{fig:domain}
\end{center}
\end{figure}

We devide the momentum space into three domains illustrated in Fig.~\ref{fig:domain},
$D_{\mathrm{in}}=\{\bm{k}\in \mathbb{R}^2:k<\varepsilon\}$,  
$D=\{\bm{k}\in \mathbb{R}^2:\varepsilon<k<R\}$, 
and $D_{\mathrm{out}}=\{\bm{k}\in \mathbb{R}^2:R<k\}$
with $\varepsilon\rightarrow 0$ and $R\rightarrow\infty$.
We consider connection between the wave functions in adjacent domains through by gauge transformation.
For $D_{\mathrm{in}}$ and $D_{\mathrm{out}}$, namely when $k=0$ and $k\rightarrow\infty$, respectively, 
the Hamiltonian $H_{+i}$ commutes with $J_z$ and $|\psi_n^{\mathrm{in/out}}\rangle=|m_z\rangle$ where $J_z |m_z\rangle=m_z |m_z\rangle$. 
Specifically, the occupied states are
\begin{align}
	m_z\in s_{\mathrm{in}} =& \left\{ \cdots \frac{7}{2}, \frac{3}{2}, -\frac{1}{2}, -\frac{5}{2} \cdots \right\} \quad \hbox{ for } D_{\mathrm{in}} \\
	m_z\in s_{\mathrm{out}} =& \left\{ \cdots \frac{5}{2}, \frac{1}{2}, -\frac{3}{2}, -\frac{7}{2} \cdots \right\} \quad \hbox{ for } D_{\mathrm{out}}.
\end{align}
Because $m_z\in s_{\mathrm{in (out)}}$, $m_z=2n-1/2$ $(m_z=2n+1/2)$ is characterized by the eigenvelue $\epsilon_k$ of the $-ie^{iJ_z\pi}\epsilon_k$ at $k=0$ 
On the other hand, for the domain $D$, 
the rotational symmetry allows us to describe the the wave function as
\begin{align}\label{eq:psin}
	|\psi_{\nu}(k,\phi_k)\rangle=\sum_{m_z} e^{i(m_z+\frac{1}{2})\phi_k} O_{\nu,m_z}(k) |m_z \rangle,
\end{align}
where $\phi_k=\arctan(k_y/k_x)$ represents the azimuthal angle and  
$O_{\nu,m_z}$ is real matrix.
For $k\rightarrow \varepsilon$ and $k\rightarrow R$ a linear combination of the occupied state
can generate the occupied angular momentum state
\begin{align}
	|\psi_{m_z}(\bm{k})\rangle
	=&\sum_{\nu\in\mathrm{occ}}\!\! O_{m_{z} ,\nu}(k) |\psi_{\nu}(k,\phi_k)\rangle\\
	=&e^{i(m_{z}+\frac{1}{2})\phi_k} |m_{z} \rangle 
\end{align}
where 
\begin{align}
	\left\{\begin{array}{l}
		m_z\in s_{\mathrm{in}} \quad \hbox{ for } k\rightarrow \varepsilon,\quad  \\
		m_z\in s_{\mathrm{out}}\quad \hbox{ for } k\rightarrow R, \quad   
	\end{array}\right.
\end{align}
Consequently, the wave function in $D_{\mathrm{in/out}}$ is described by the $U(1)$ gauge transformation from those in domain $D$,
\begin{align}
	|\psi_{m_z}^{(\mathrm{in/out})} \rangle = e^{-i\lambda^{(\mathrm{in/out})}_{m_z}} |\psi_{m_z} \rangle.
\end{align}
with
\begin{align}\label{eq:chinow}
	\lambda_{m_{z}}^{\mathrm{(in/out)}} = \left(m_{z} + \frac{1}{2}\right)\phi_k
\end{align}
for $m_z\in s_{\mathrm{in/out}}$.
In addition, the Berry connection in $D_{\mathrm{in/out}}$ is given by
\begin{align}\label{eq:gtrans}
	\bm{a}^{(\mathrm{in/out})}_{m_z} = \bm{a}_{m_z} + \bm{\nabla}_{\bm{k}}\lambda^{\mathrm{in/out}}_{m_z}
\end{align}
Finally, using the gauge transformation Eq.~\eqref{eq:gtrans}, the Chern number is determined as
\begin{align}
	C_+=&\sum_{\nu\in \mathrm{occ}} \Bigg[\int_{D_{\mathrm{in}}} \frac{dk}{2\pi} \left(\bm{\nabla}_{\bm{k} }\times \bm{a}_{\nu}^{\mathrm{in}}\right)_z 
	+ \int_{D} \frac{dk}{2\pi} \left(\bm{\nabla}_{\bm{k} }\times \bm{a}_{\nu}\right)_z\nn\\
	&+ \int_{D_{\mathrm{out}}} \frac{dk}{2\pi} \left(\bm{\nabla}_{\bm{k} }\times \bm{a}_{\nu}^{\mathrm{out}}\right)_z\Bigg] \nn\\
	=&\frac{1}{2\pi}\sum_{\nu\in \mathrm{occ}} \Bigg[ \int_{\partial_\mathrm{in}} d\bm{k}\cdot \bm{\nabla}_{\bm{k}}\lambda^{\mathrm{in}}_{\nu} 
	+ \int_{-\partial_\mathrm{out}}\!\!\! d\bm{k}\cdot \bm{\nabla}_{\bm{k}}\lambda^{\mathrm{out}}_{\nu}\Bigg] \nn\\
	=&\sum_{m_z\in s_\mathrm{in}} \left(m_z+\frac{1}{2}\right) - \sum_{m_z\in s_\mathrm{out}} \left(m_z+\frac{1}{2}\right) \nn\\
	=&1-3+5-7 + \cdots+(-1)^{J+\frac{3}{2}}2J\nn \\ 
	=& \pm \left(J+\frac{1}{2}\right) \label{eq:cgauge}
\end{align}
In the second equality, we have useed the Stokes theorem to 
change the area integral of $D_{\mathrm{in/out}}$, and $D$ to line integral along $\partial_{\mathrm{in}}$ and $-\partial_{\mathrm{out}}$ shown in Fig.~\ref{fig:domain}.
The double sign in the final result Eq.~\eqref{eq:cgauge} 
applies to the cases when $J=2n+\frac{1}{2}$ (e.g.,$J=\frac{1}{2}$, $\frac{5}{2}$, etc.) and $J=2n+\frac{3}{2}$ (e.g.,$J=\frac{3}{2}$, $\frac{7}{2}$,etc.), respectively.

\begin{figure*}[t]
\begin{center}
	\includegraphics[width=170mm]{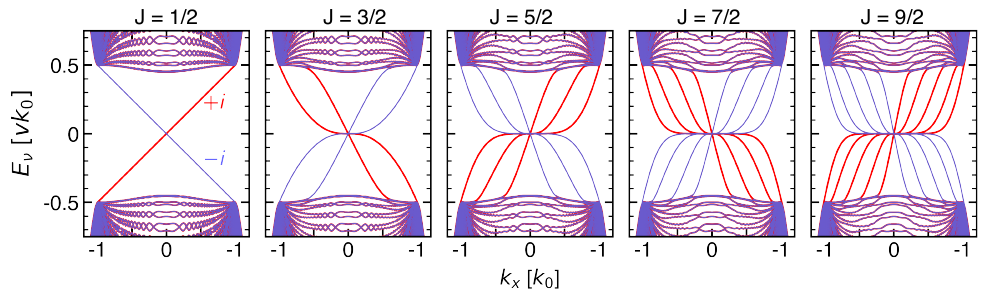}
	\caption{The same plot as Fig.~\ref{fig:spectrum}(a), but with an extended range of spins, up to $J=9/2$.
	Notably, as the spin $J$ increases, the in-gap states within the mirror $\pm i$ 
	exhibit a systematic grouth in the number of branches, a phnomenon elucidates by 
	the mirror Chern numbers discussed in Eq.~\eqref{eq:cmmain} in the main text.
	}\label{fig:1to9}
\end{center}
\end{figure*}

It is clear that the Chern number 
in the sector characterized by mirror eigenvalue $\eta_{M_z}=-i$ is given by
\begin{eqnarray}
	C_{-} = -C_{+}
\end{eqnarray}
The mirror Chern number, which represents the difference in Chern number between $\eta_{M_z}=\pm i$ sectors 
is expressed as:
\begin{eqnarray}
	C_{M_z}=\frac{1}{2}(C_{+}  -C_{-} ) = \pm (J+\frac{1}{2}).
\end{eqnarray}
This corresponds to Eq.~\eqref{eq:cmmain} in the main text.
To provide clearler insight, in Fig.~\ref{fig:1to9}, 
we present the energy spectrum, involving a wider range of spins, with values up to $J=9/2$.
It illustrates the bulk-edge correspondence associated with the mirror Chern number, 
as discussed in the main text.

\section{Derivation of effective Hamiltonian} \label{sec:perturb}
We derive the effective Hamiltonian given by Eq.~\eqref{eq:heff} 
using perturbation theory. 
Our unperturbed Hamiltonian is described in Eq.~\eqref{eq:h0}, 
consisting of plane-wave states Eq.~\eqref{eq:pw} 
and degenerate zero-energy states localized at edge Eq.~\eqref{eq:edge}.
For sufficiently small $k_x$ and $k_x \neq 0$, 
we expand the 0-th order edge state as 
\begin{align}
	| u_\beta \rangle = \sum_{m} u_{\beta, m} | \chi_{m} \rangle
\end{align}
in the basis of zero energy states in Eq.~\eqref{eq:edge}.
Within this framework, we analyze the Hamiltonian $H_{\mathrm{eff}}(k_x)$, 
which effectively lifts the degenracy of $| \chi_{m} \rangle$ states
and simultaneously determines $u_{\beta,m}(k_x)$.

Assuming small value of $k_x$, 
and following conventions of perturbation theory, 
we invoke a perturbation series to describe both the energy and the wave function
\begin{gather}
	\mathcal{E}_{\beta} (k_x) = \sum_{n=1} \left(\frac{k_x}{k_0}\right)^{n} E^{(n)}_{\beta} \\ 
	| \psi_{\beta}(k_x) \rangle = |u_\beta \rangle + \sum_{n=1} \left(\frac{k_x}{k_0}\right)^{n} | u_{\beta} ^{(n)} \rangle.
\end{gather}
In these expressions it should be noted that the energy of the 0-th order is zero, leading to $H_0 |u_{\beta}\rangle=0$. 

To satisfy the Schr\"odinger equation $H(k_x)| \psi_{\beta}(k_x) \rangle = \mathcal{E}_{\beta}(k_x)| \psi_{\beta}(k_x)\rangle$
in each order of $k_x^{n}$, a series of equations emerges:
\begin{align}
	&H_{0} | u_{\beta}^{(1)}\rangle + (H_1-E_{\beta}^{(1)}) |u_{\beta} \rangle = 0 \label{eq:p1}\\
	&H_{0} | u_{\beta}^{(2)} \rangle + (H_1-E_{\beta}^{(1)}) |u_{\beta}^{(1)}\rangle = E_{\beta}^{(2)} | u_{\beta} \rangle \label{eq:p2}\\
	&H_{0} | u_{\beta}^{(3)} \rangle + (H_1-E_{\beta}^{(1)}) |u_{\beta}^{(2)}\rangle = E_{\beta}^{(2)} | u_{\beta}^{(1)} \rangle +E_{\beta}^{(3)} | u_{\beta} \rangle  \label{eq:p3}\\
	& \quad \quad \quad \vdots \nn \\
	&H_{0} | u_{\beta}^{(n)} \rangle + (H_1-E_{\beta}^{(1)}) |u_{\beta}^{(n-1)}\rangle \nn \\
	& =  E_{\beta}^{(2)} | u_{\beta}^{(n-2)} \rangle +E_{\beta}^{(3)} | u_{\beta}^{(n-3)} \rangle + \cdots + E_{\beta}^{(n)} | u_{\beta} \rangle \label{eq:pn}
\end{align}
In addition, perturbation ket vector is expanded by plane wave states,
\begin{align}
	| u_\beta^{(n)}\rangle = \sum_{\nu} u_{\beta; k_y,m,\eta}^{(n)} | \nu \rangle.
\end{align}
in the following procedure. See also Eq.~\eqref{eq:pw} in the main text for the definition of the plane-wave states, with joint index $\nu=(k_y,m,\eta)$. 

We sequentially analyze the perturbation equations~\eqref{eq:p1}, \eqref{eq:p2}, \eqref{eq:p3}, \eqref{eq:pn}.
When we apply $\langle \chi_{m_y}|$ to Eq.~\eqref{eq:p1}, 
we drive an eigenvalue equation:
\begin{eqnarray}\label{eq:ev1}
	\sum_{m'} \langle \chi_{m_y}|H_1|\chi_{m'} \rangle u_{\beta, m'}= E_{{\beta}}^{(1)} u_{\beta, m}.
\end{eqnarray}
It is worth noting that $\langle \chi_{m}|H_1|\chi_{m'}\rangle (k_x/k_0)$ corresponds to the first-order term of effective Hamiltonian $H_{\mathrm{eff}}$.

To narrow down the possibilities for the matrix elements of  $\langle \chi_{m}|H_1|\chi_{m'}\rangle$ and 
provide a crucial insight into the nature of the perturbation within the present system
we utilize two convenient selection rules applicable to any order of perturbation. 
The first rule arises from the chiral symmetry inherent in the zero-field Hamiltonian $H$ given in Eq.~\eqref{eq:hbulk}: 
\begin{align}
	\gamma H(k_x) \gamma^{\dag} = -H(k_x) \quad \hbox{with }  \gamma=\sigma_y.
\end{align}
In the subspace 
\begin{align}
	S=\{|\chi_{J}\rangle, \cdots,|\chi_{\frac{1}{2}}\rangle; | \chi_{-\frac{1}{2}}\rangle, \cdots ,|\chi_{-J}\rangle\},
\end{align}
the chiral operator takes a diagonal form:
\begin{align}
	\langle \chi_{m} | \gamma | \chi_{m'} \rangle = \delta_{m,m'} \sgn(m). 
\end{align}
Consequently, the effective Hamiltonian with chiral symmetry $\gamma H_{\mathrm{eff}}\gamma^\dag=-H_{\mathrm{eff}}$ 
should adopt an off-diagonal block form: 
\begin{align}\label{eq:block}
	H_{\mathrm{eff}}=
	\left(\begin{array}{cc}
	0 & D \\
	D^{\dag} & 0
	\end{array}\right).
\end{align}
in the basis of $S$.
This insight indicates that the effective Hamiltonian, at any order of perturbation stemming from zero-field Hamiltonian,
avoids coupling states $|\chi_{m}\rangle$ with opposite signed angular momenta, specifically, $m>0$ and $m<0$.

The second selection rule stems from the conservation of the angular momentum. 
Given that the perturbation $H_1=v J_x\sigma_x$ corresponds to the first order of $J_x$, 
it can couple angular momentum states $m_y$ and $m_y'=m_y\pm1$.
By combining these two aforementioned rules, we deduce that the $n$-th order perturbation term exhibits 
non-zero matrix elements within the subspace $s=\{ |\chi_{n/2}\rangle, |\chi_{n/2-1}\rangle, \cdots, |\chi_{-n/2}\rangle \}$.

Specifically, non-zero matrix elements of first order perturbation described by Eq.~\eqref{eq:ev1} appear only in the subspace $s_{1/2}=\{|\chi_{1/2}\rangle, |\chi_{-1/2}\rangle\}$  
with $h_{1/2}=\langle \chi_{-{1}/{2}}|H_1|\chi_{{1}/{2}}\rangle$,
resulting in the formulation of the effective Hamiltonian as follows:
\begin{align}
	H_{\mathrm{eff}}^{(1/2)}=
	\left(\begin{array}{cc}
		0 & h_{1/2}^\ast k_x/k_0 \\
		h_{1/2} k_x/k_0 & 0 \\
	\end{array}\right).
\end{align}
Notably, this effective Hamiltonian corresponds to the massless one-dimensional Dirac Hamiltonian. 
Here, the matrix element can be conveniently calculated:
\begin{align}
	h_{1/2} = -\frac{iv k_0}{2}  \left(J+\frac{1}{2}\right).
\end{align}

On the other hand, the higher spin states $|m| \ge \frac{3}{2}$ remain degenerate at $E^{(1)}_{\beta}=0$ for the first order perturbation. 
To investigate the eventual splitting of these states, it becomes necessary to explore the higher-order perturbation in the complementary space 
\begin{align}\label{eq:s1}
	S_{3/2}=\{|\chi_{J}\rangle, \cdots, |\chi_{3/2}\rangle ; |\chi_{-3/2}\rangle,\cdots |\chi_{-J}\rangle\}.
\end{align}
By applying the plane wave state $\langle\nu|=\langle k_y, m, \eta|$ to Eq.~\eqref{eq:p1},
we acquire
\begin{align}\label{eq:g1}
	u_{\beta; k_y,m_y,\eta}^{(1)} = -\sum_{m'}\frac{\langle k_y, m, \eta|H_1|\chi_{m'}\rangle}{E_{k_y,m,\eta}} u_{\beta,m'}
\end{align}
Upon substituting Eq.~\eqref{eq:g1} and subsequently applying $\langle \chi_{m}|$ to Eq.~\eqref{eq:p2},
we derive an eigenvalue equation 
\begin{align}\label{eq:ev2}
	-\!\!\!\!\sum_{m'\in S_{3/2}}\!\!\!\!  \langle \chi_{m_y}| H_1\sum_{k_y,m''}P_{m''}(k_y)H_1  |\chi_{m'} \rangle u_{\beta, m'} 
	= E_{{\beta}}^{(2)} u_{\beta, m},
\end{align}
where we introduce the projection operator
\begin{align}
P_{m}(k_y) = \sum_{\eta}  \frac{|k_y, m, \eta\rangle\langle k_y, m, \eta |}{{ E_{k_y, m, \eta}}}.
\end{align}
It is important to note that due to the selection rule associated with the chiral symmetry Eq.~\eqref{eq:block} 
and the conservation of angular momentum, 
all second-order matrix element within Eq.~\eqref{eq:ev2} for subspace $S_{3/2}$ are found to be zero.

Furthermore, within the exact same procedure, 
we can readily derive the eigenvalue equation for the third-order perturbation energy: 
\begin{align}
	\sum_{m'\in S_{3/2}}  \langle \chi_{m}| H_1\left(\sum_{k_y,m''}P_{m''}(k_y)H_1 \right)^2 |\chi_{m'} \rangle u_{\beta, m'} \nn\\
	= E_{{\beta}}^{(3)} u_{\beta, m}.
\end{align}
This analysis highlights the lifting of degeneracy in the subspace $s_{3/2}=\{|\chi_{3/2}\rangle, |\chi_{-3/2}\rangle\}$ 
induced by the third-order perturbation. 
This process results in the formulation of an effective Hamiltonian with the following form:
\begin{align}\label{eq:h32}
	H_{\mathrm{eff}}^{(3/2)}=
	\left(\begin{array}{cc}
		0 & h_{3/2}^\ast(k_x/k_0)^3 \\
		h_{3/2}(k_x/k_0)^3 & 0 \\
	\end{array}\right).
\end{align}
An important feature of this results arises from the absence of lower-order perturbation energy $E^{(1)}_{\chi_\beta} =E^{(2)}_{\chi_\beta}=0$
within the considered subspace $\beta\in S_{3/2}$.
This absence contributes to the systematic derivation of ~\eqref{eq:h32} and systematic formulation. 

In the same manner, perturbation up to $2m$-th order 
results in the effective Hamiltonian for the subspace $s_{|m|}=\{|\chi_{|m|}\rangle, |\chi_{-|m|}\rangle\}$
\begin{align}
	H_{\mathrm{eff}}^{(|m|)}=
	\left(\begin{array}{cc}
		0 & h_{|m|}^\ast(k_x/k_0)^{|2m|} \\
		h_{|m|}(k_x/k_0)^{|2m|} & 0 \\
	\end{array}\right).
\end{align}
associated with a matrix element, 
\begin{align}
	h_{|m|} = \langle \chi_{-|m|}| H_1\left(\sum_{k_y,m''}  P_{m''}(k_y) H_1 \right)^{2m-1} |\chi_{|m|} \rangle.
\end{align}
These expressions correspond to Eq.~\eqref{eq:heff} with Eq.~\eqref{eq:coeff}.

This systematic exploration of higher order perturbations further 
strengthens our understanding of interplay between perturbations and distinct behavior of the 
higher spin states in the present system.

\section{Derivation of non-linear conductivity} \label{sec:spinmat}
Here we present the derivation of the linear and non-linear conductivity as given by the Eq.~\eqref{eq:sigma}.
The effective Hamiltonian Eq.~\eqref{eq:heff} provides 
the dispersion relation 
\begin{equation}
	E_{\pm,|m|,k_x} = \pm h_{|m|} k_x^{|2m|}
\end{equation}
By sequentially substituting Eq.~\eqref{eq:boltzman} into itself, we obtain a formal solution of the non-equilibrium distribution function
\begin{align}
	f_{k_x}=f_0 + \left(-\frac{\mathcal{E}\tau}{\hbar}\right)\frac{\partial f_0(E_{\pm,|m|,k_x})}{\partial k_x} \nn\\ 
	+ \left(-\frac{\mathcal{E}\tau}{\hbar}\right)^2\frac{\partial^2 f_0(E_{\pm,|m|,k_x})}{\partial k_x^2} + \cdots
\end{align}
Consequently, contribution to non-equilibrium current 
from the band labeled by $\pm$ and $|m|$ can be expressed as:
\begin{widetext} 
\begin{align}\label{eq:jpmmy}
	j_{\pm,|m|} &= \frac{1}{L}\sum_{k_x}\Big(f_{k_x}+f_{0}(E_{\pm,|m|,k_x})\Big)\frac{1}{\hbar}\frac{\partial E_{\pm,|m|,k_x}}{\partial k_x}\nn\\
	 &= \frac{1}{2\pi\hbar}\int_{-\infty}^{\infty} dk_x\Big(f_{k_x}+f_{0}(E_{\pm,|m|,k_x})\Big) \frac{\partial E_{\pm,|m|,k_x}}{\partial k_x} \nn\\
	 &= \frac{1}{2\pi\hbar}\sum_n \int_{-\infty}^{\infty} dk_x\left(-\frac{\mathcal{E}\tau}{\hbar}\right)^n \frac{\partial^nf_0( E_{\pm,|m|,k_x})}{\partial k_x^n} \frac{\partial E_{\pm,|m|,k_x}}{\partial k_x} \nn\\
	&= -\frac{1}{2\pi\hbar}\sum_n \int_{-\infty}^{\infty} dk_x\left(\frac{\mathcal{E}\tau}{\hbar}\right)^n \frac{\partial f_0( E_{\pm,|m|,k_x})}{\partial k_x} \frac{\partial^n E_{\pm,|m|,k_x}}{\partial k_x^n}\nn\\
	&= -\mathrm{sign}(\pm h_{|m|})\frac{1}{2\pi\hbar}\sum_n \int_{-\infty}^{\infty} dE_{\pm,|m|,k_x}\frac{\partial k_x}{\partial E_{\pm,|m|,k_x}}\left(\frac{\mathcal{E}\tau}{\hbar}\right)^n \frac{\partial f_0( E_{\pm,|m|,k_x})}{\partial E_{\pm,|m|,k_x}} \frac{\partial E_{\pm,|m|,k_x}}{\partial k_x}\frac{\partial^n E_{\pm,|m|,k_x}}{\partial k_x^n}\nn\\
	&= \mathrm{sign}(\pm h_{|m|})\frac{1}{2\pi\hbar}\sum_n \left(\frac{\mathcal{E}\tau}{\hbar}\right)^n \left(\frac{\partial^n E_{\pm,|m|,k_x}}{\partial k_x^n} \right)_{k_x=k_F}\nn\\
	&=\frac{|h_{|m|}|}{2\pi\hbar}\left(\frac{\mathcal{E}\tau (|2m|)!}{\hbar}\right)^{|2m|}
\end{align}
\end{widetext} 
For the second identity, we replace $\frac{1}{\delta k} \sum_{k}\delta k$ with $\frac{1}{\delta k}\int d k$ using $\delta k = 2\pi/L$.
In the case of the fourth dentity we apply integration by parts $n-1$ times.
The fifth identity involves a change of integration variables from wave number $k_x$ to energy $E$. 
The sixth identity relies on $\frac{\partial f_0(E)}{\partial E}=-\delta(E-E_F)$ for the zero temperature limit.
The last identity is derived from $k_F=0$ in the present system

Finally, combining Eq.~\eqref{eq:jpmmy} with the definition
\begin{align}
	j=\sum_{\pm, |m|} j_{\pm, |m|} = \sum_n \sigma^{(n)} \mathcal{E}^n, 
\end{align}
we obtain 
\begin{align}
\sigma^{(|2m|)} =  \frac{\tau^{|2m|} (|2m|)!}{\pi \hbar^{|2m|+1}}  |h_{|m|}|
\end{align}
This completes the derivation process, providing the linear and non-linear conductivity in the context of the system described by the present effective Hamiltonian.


\bibliography{reference}

\end{document}